\documentclass{svproc}
\usepackage{url}
\usepackage{graphicx}

\usepackage{hyperref}

\begin{document}
\mainmatter              
\title{A Tool for Automatic Estimation of Patient Position in Spinal CT Data}
\titlerunning{Patient position Detection}  
%
\author{Roman Jakubicek\inst{1} \and Tomas Vicar\inst{1} \and Jiri Chmelik\inst{1} }
\authorrunning{Roman Jakubicek et al.} 
%
\tocauthor{Roman Jakubicek, Tomas Vicar and Jiri Chmelik}
\institute{Department of Biomedical Engineering, Faculty of Electrical Engineering and Communication, Brno University of Technology, Brno, CZE\\
\email{jakubicek@vutbr.cz},\\ 
WWW home page:
\texttt{\url{https://www.vutbr.cz/en/people/roman-jakubicek-119710}}\newline
\textbf{preprint - submitted on EMBEC 2020 - paper has not been reviewed yet}
}

\maketitle              

\begin{abstract}
Much of the recently available research and challenge data lack the meta-data containing any information about the patient position. This paper presents a tool for automatic rotation of CT data into a standardized (HFS) patient position. The proposed method is based on the prediction of rotation angle with CNN, and it achieved nearly perfect results with an accuracy of 99.55 \%. We provide implementations with easy to use an example for both Matlab and Python (PyTorch), which can be used, for example, for automatic rotation correction of VerSe2020 challenge data.

\end{abstract}
\section{Introduction}
Using computer methods for medical image analysis has become a standard supportive tool of a medical diagnosis. Many methods require a specific patient position in the image data for its proper running. Generally, the most usual position is HFS (Head First Supine), but it depends on conducted examination (and other acquisition parameters and scanning options). For example, the HFS patient position is usually used for examination of a head/brain, lung scans, whole-body scans, spine scans, etc. On the other hand, a prone position can be used during the colonoscopy, and feet first position during the angiography of lower limbs or abdomen examination.

In most cases, the exact patient position in the scanner (and in image data) is stored in a meta-data file or a header of the image file. It can be easily retrieved and used in a pre-processing step (geometric transformation of image data). A form of patient position information depends on the image file format (e.g., DICOM, Nifty, MHD/RAW). In those cases, the automatic estimation of the patient position is not important. However, much of the recently available research and challenge data are lack of the meta-data containing any information about the patient position. Here, the patient position can be manually or semi-automatically identified during the pre-processing of the data before the main analysis, which is too time-consuming, especially if there are hundreds of data. In this paper, we are working with data from VerSe2020 challenge database \cite{Loeffler2020,Sekuboyina2020,Sekuboyina2020verse} and own databases, but several other public databases exist.

Due to this, we offer a fully automatic solution for the estimation of the patient position in 3D CT data. Nowadays, deep learning algorithms are state-of-the-art methods in medical image processing tasks \cite{Litjens2017,Shen20170621,Ker2018}. Our algorithm uses a convolution neural network, and its source code is publicly available in two versions(2D and 3D CNN). The code can be downloaded from GitHub for Python and Matlab (\url{https://github.com/JakubicekRoman/CTDeepRot}).

\section{Methods}
For the rotation angle's estimation, enabling the transformation of CT data into the HFS patient position, the CNNs have been trained in two versions - with 2D projections and original 3D volume.

\subsection{Rotation prediction}

There are 24 possible unique 3D rotations with 90 deg steps (rotation group of the cube has 24 elements). List of all possible rotations can be, for example, created by combinations of rotation around x-axis $a$, y-axis $b$ and z-axis $c$ as follows: \{$a^2,a^2c,a^2c^2,a^2c^3,a^2b^2,a^2b^2c,a^2b^2c^2,a^3c,a^3c^2,a^3c^3,a^3b,a^3bc,a^3bc^2,a^3bc^3,a^3b^2,$\\ $a^3b^2c,a^3b^2c^2,a^3b^2c^3,a^3b^3,a^3b^3c,a^3b^3c^2,a^3b^3c^3$\}, where, for example, $a^2c$ is 180 deg rotation around x-axis followed by 90 deg rotation around the z-axis. 

\begin{figure}[!tb]
\centering
\includegraphics[width=0.95\linewidth]{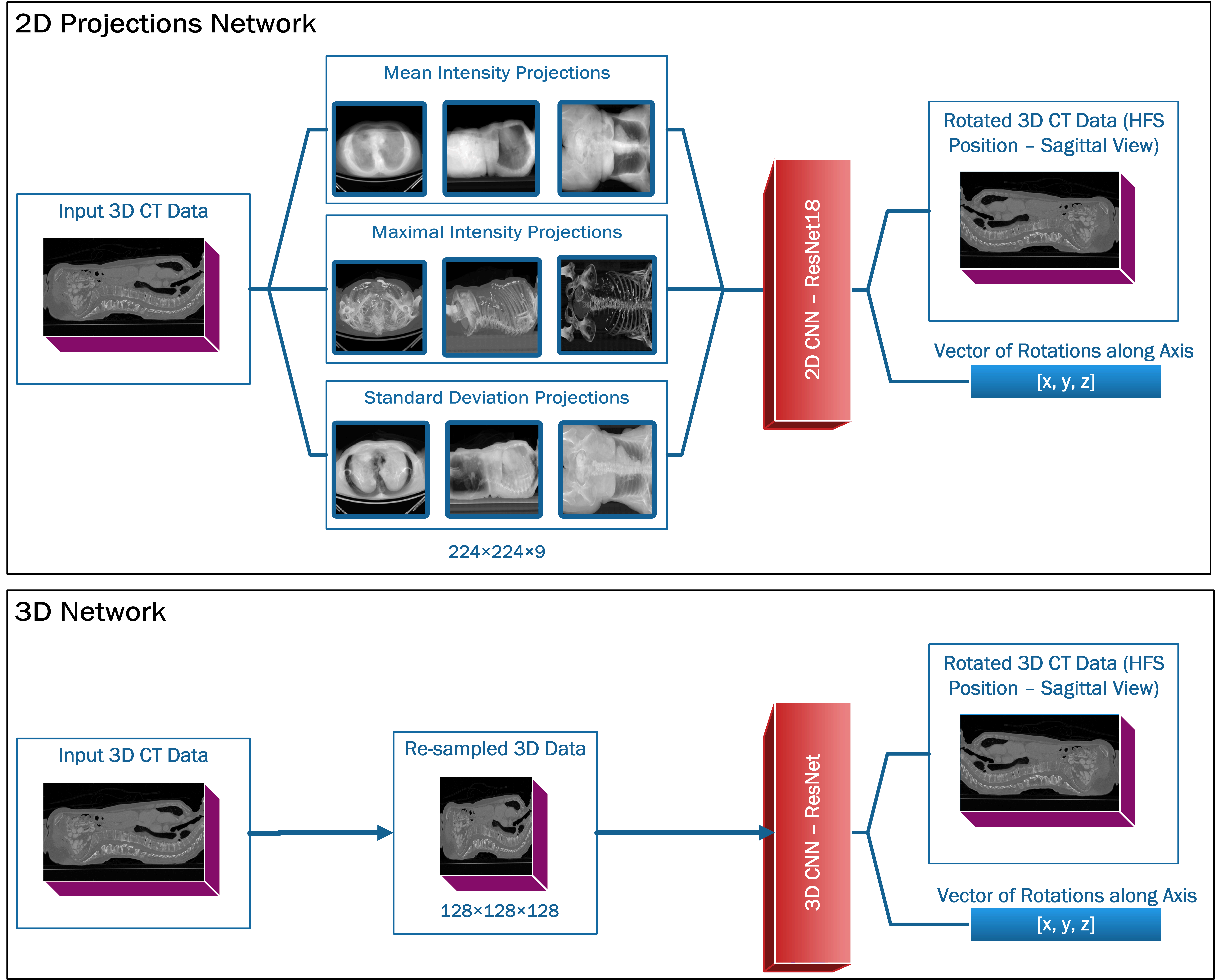}
\caption{Block scheme of the 2D and 3D versions of the proposed approach.}
\label{fig:block_scheme}
\end{figure}

In this work, rotation angles estimation is formulated as a classification problem into 24 classes, which correspond to 24 unique patient positions. Rotation angles are assigned to each category, which can be subsequently used to transformation of data into the HFS position.
This task's solution is based on the angle estimation via a convolution neural network with ResNet architecture \cite{resnet18}. The last fully-connected layer classifies each input data into 24 classes. 
In the case of 2D projections, input data are represented as a 9-channel 2D image consisting of Maximum Intensity Projections (MIP) in three views (in the direction axis of X, Y, and Z), three mean projections and three ''projection'' of standard deviations. For version 3D, res-sampled original data are taken. Block schemes of rotational angle prediction for 2D and 3D approaches are shown in Fig. \ref{fig:block_scheme}.



\subsection{Implementation details}


Both 2D and 3D networks were trained using Matlab and Python (PyTorch), where batch-size was 32 and 8 for 2D and 3D networks. All networks were trained using Adam \cite{Kingma2015AdamAM} optimizer with an initial learning rate of 0.001 ($\beta_1=0.9$ and $\beta_2=0.99$) for 14 epochs; the learning rate was reduced to 10\% every six epochs. Input images were resized to the size $224\times224$ and $128\times128\times128$ for 2D and 3D networks, respectively. Input volume/each projection type was normalized from manually defined range to range $<$-0.5, 0.5$>$. In the 2D case, ResNet-18 \cite{resnet18} architecture was used. For the 3D case, the 3D equivalent of ResNet-18 was designed, where all 3D operations were replaced by its 3D equivalents with a smaller number of convolution filters.

\subsection{Experimental data}
The used dataset contains 492 CT scans from various databases and with other scanning protocols. However, we focus primarily on spinal CT data. This database includes publicly available datasets (VerSe2019 and VerSe2020 \cite{Loeffler2020,Sekuboyina2020,Sekuboyina2020verse}, some data from SpineWeb\footnote{available from \url{http://spineweb.digitalimaginggroup.ca/spineweb/index.php?action=home}} \cite{glocker2013vertebrae,yao2016multi}) and own data from several cooperating medical institutions. There are many scanned patients in an advanced stage of cancer with deformed vertebrae and scoliotic spine, and they often have surgical implants. For all CT scans, the ground-truth rotation was manually defined, and all 24 possible rotations were generated and used for training/evaluation.

\begin{figure}[!tb]
\centering
\includegraphics[width=0.95\linewidth]{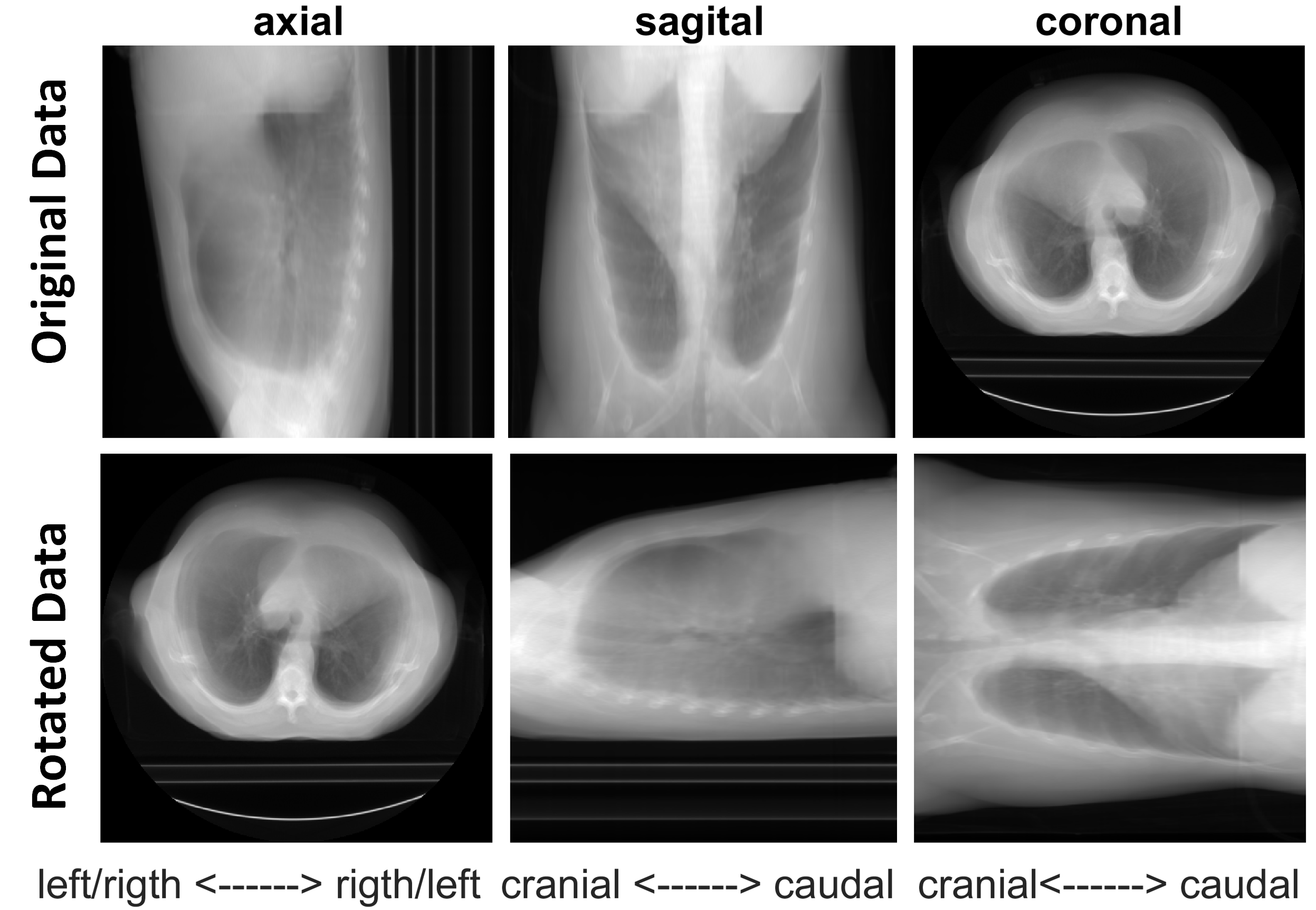}
\caption{Example of original data and data rotated by CNN predicted angles -- mean intensity projections. First row -- original uncorrected data, second-row -- rotated data into Head First Supine (HFS) patient position.}
\label{fig:prediction_example}
\end{figure}

%

%
\section{Results and Discussion}
Twenty percent of scans, randomly selected from all scans, were used to test the models and remaining data for their training. 
Achieved prediction accuracy with the 2D network was higher 99.55 \% compared to 98.63 \% accuracy of the 3D network.

\begin{figure}[!tb]
\centering
\includegraphics[width=0.95\linewidth]{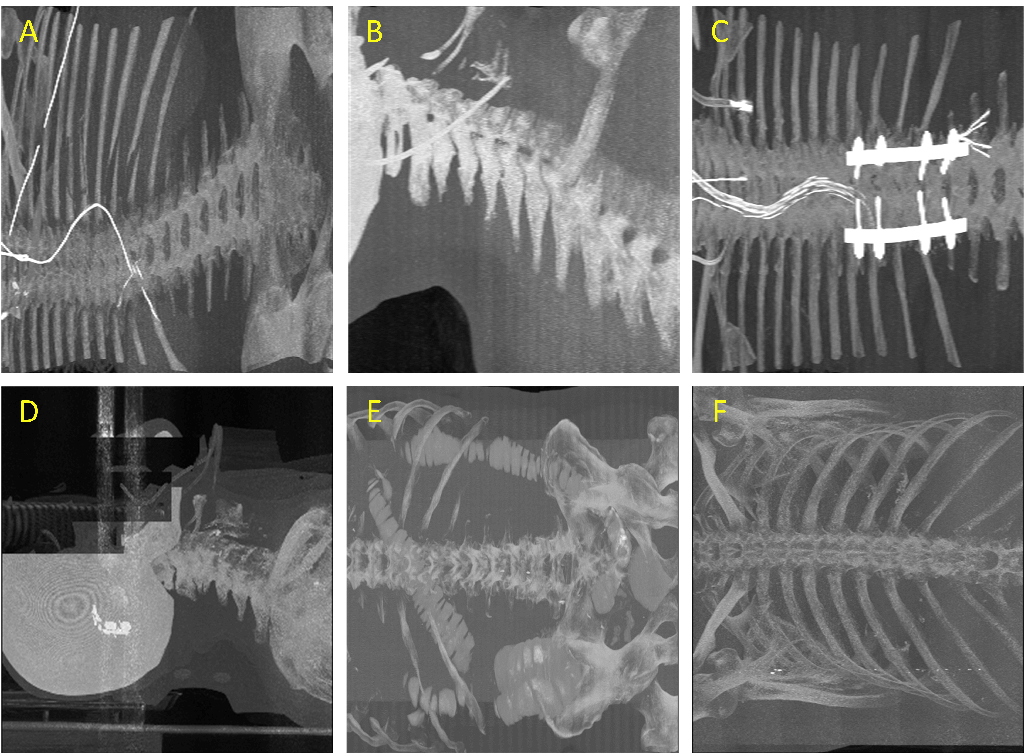}
\caption{Example of images where rotation was not correctly predicted -- maximal intensity projections. A, B, C -- scoliotic spine with surgical implants and catheters, D -- data with removed parts, E -- colonoscopic image with contrast media in the colon, F -- short scan of lungs.}
\label{fig:prediction_example_failed}
\end{figure}

Both networks (2D and 3D) achieved very good accuracy, making mistakes in only very complicated cases. One such example can be an occurrence of surgical implants, which distinctly affect contrast resolution (especially in 2D version using global projections), and it can lead to a wrong prediction of the angle (Fig.~\ref{fig:prediction_example_failed}~A-C). Further, based on validation, it was found that often the reason for mistakes was too short scans (along Z-axis of the patient), e.g., a local lung scan only in the thoracic area (Fig.~\ref{fig:prediction_example_failed}~F). In most cases of these wrong predictions, the algorithm provides rotated 3D data, which are falsely rotated into the FFP patient position.

In the clinical medicine for CT examination, there are eight official patient positions. Our proposed network is able to detect all these positions. The remaining rotation angles, which provide the non-real positions, may occur in especially research databases (such as VerSe2020 \cite{Sekuboyina2020verse}) as a result of the transfer, re-writing, or desirable geometrical transformations.

Publicly available database for mentioned challenge VerSe2020 \cite{Sekuboyina2020verse} contains various rotated CT data primarily focusing on vertebra segmentation. Our algorithm can be used as a pre-processing step for this challenge, especially if research teams' algorithms for segmentation require the standard patient position. In this case, the alignment of data can be done manually (taking a few minutes per scans) or automatically rotated via the proposed tool. It should be emphasized that this database contains also data with non-real patient positions. An example of a correctly rotated scan is shown in Fig.~\ref{fig:prediction_example}.


\section{Conclusion}

This paper presents a tool for automatic rotation of CT data into a standardized (HFS) patient position. The proposed method is based on the prediction of rotation angle with CNN, and it achieved nearly perfect results with an accuracy of 99.55 \% and in a very short time (in the order of units of seconds per scan). We provide implementations with easy to use an example for both Matlab and Python (PyTorch), which can be used, for example, for automatic rotation correction of VerSe2020 \cite{Sekuboyina2020verse} challenge data.

\section*{Acknowledgement}
The Titan Xp GPU used for this research was donated by the NVIDIA Corporation. We acknowledge the contribution of the Osteo-Oncology Center, Istituto Scientifico Romagnolo per lo Studio e la Cura dei Tumori (I.R.S.T.) S.r.l., Meldola (Italy) and St. Anne's University Hospital, Brno (Czech Republic), who provided part of CT image data.

\section*{Conflict of Interest}
Authors declare none.

%
%
\bibliographystyle{splncs03}
\bibliography{mybibfile}

\end{document}